\documentstyle[preprint,tighten,aps,epsfig]{revtex}

\newcommand{\etal}{\mbox{\em et al}\,}
\newcommand{\sw}{\sin\theta_W}
\newcommand{\cw}{\cos\theta_W}
\newcommand{\im}{\mbox{Im}}
\newcommand{\re}{\mbox{Re}}
\newcommand{\swc}{\sin^2\theta_W}
\newcommand{\beq}{\begin{equation}}
\newcommand{\eeq}{\end{equation}}
\newcommand{\mco}{\multicolumn}
\begin{document}

\preprint{
\begin{tabular}{r} FTUV/99$-$1 \\ IFIC/99$-$1
\end{tabular}
}

\title{Electric and Weak Electric  Dipole Form Factors for Heavy Fermions
in a 
General Two Higgs Doublet Model}
\author{D.\ G\'omez Dumm}
\address{\hfill \\ Departament de F\'{\i}sica Te\`orica, IFIC, 
CSIC -- Universitat de Val\`encia \\
Dr.\ Moliner 50, E-46100 Burjassot (Val\`encia), Spain}
\author{G.A. Gonz\'alez-Sprinberg}
\address{\hfill \\ Instituto de F\'\i sica, Facultad de Ciencias, 
Universidad de la Rep\'ublica\\ 
Igu\'a 4225, Montevideo 11400, Uruguay}

\maketitle

\begin{abstract}
The electric and weak electric dipole form factors for heavy fermions
are calculated in the context of the most general two-Higgs-doublet model
(2HDM). We find that the large top mass can produce a significant
enhancement of the electric dipole form factor in the case of the $b$ and
$c$ quarks. This effect can be used to distinguish between different
2HDM scenarios.
\end{abstract}




\section{Introduction}

One of the simplest extensions of the Standard electroweak model (SM)
is the so-called two-Higgs-doublet model (2HDM), in which the new
ingredient is the presence of a second doublet of scalar fields. The
inclusion of this new fields implies various phenomenological
consequences, which have led the 2HDM to be subject of analysis during
the last two decades.

We concentrate here on one of the most interesting features concerning
the 2HDM, which is the presence of different sources of CP violation
beyond the standard $\delta_{CKM}$ phase in the quark--mixing matrix. In
particular, we analyse the effects of the new parameters of the model on
the CP--violating electric and weak electric dipole form factors for heavy
fermions. The interest on these observables has
increased in recent years, in view of the ongoing activity both in the 
theoretical and experimental areas\cite{ale,nos1,exp}. SM predictions
for CP--odd dipole moments are extremely small, and this opens
the possibility for one--loop effects coming from extended models to
show up~\cite{bern,susy}. Specific observables have been proposed and
studied in the literature\cite{ale,nos1,nel}, and some bounds 
have already been obtained from experimental measurements in
$e^+ e^-$ collisions~\cite{exp,sant}.

If done in a completely general way, the addition of a second scalar
doublet to the SM Lagrangian is problematic: one immediately finds that
a general 2HDM model contains
tree--level flavour--changing neutral currents (FCNC), which are strongly
suppressed phenomenologically. To avoid this problem, it is usual to
introduce {\em ad hoc} discrete symmetries, in such a way that
all fermions of a given charge couple to only one of the doublets
\cite{glawei}. This can be done in different ways, leading to
so-called 2HDM I and II. It is often said that the obtained flavour
conservation is ``natural''. The inclusion of discrete symmetries,
however, is not the only way
of preventing the undesired FCNC \cite{mod,cs,afs}. In fact, the
presence of
strong hierarchies in the fermion masses and mixing angles seems to be
a clear signature of an underlying theory of flavour beyond the
SM Yukawa couplings. From this point of view, it can be also ``natural''
to expect that the suppression of FCNC observed at low energies could
be explained in the context of this by now unknown theory. On the other
hand, whereas
the phenomenological constraints on FCNC are very stringent for processes
which involve the first family of quarks and leptons, this is not
the case if one considers only the mixing between the second and third
fermion families. One possibility is to assume that the suppression of FCNC
is related to the masses of the involved fermions, as has been proposed
by several authors in the literature \cite{cs,afs}.

Here, instead of choosing a particular {\it Ansatz} to enforce the suppression
of tree--level FCNC, we will consider a completely general 2HDM, using
a convenient parametrisation to take into account the existing
phenomenological constraints. As has been pointed out recently \cite{wolf},
the various sources of CP violation can be classified into four classes:
\begin{itemize}
\item CP violation in charged and neutral flavour--conserving
scalar exchange
\item CP violation in neutral flavour--changing scalar
exchange
\item CP violation in the neutral scalar mixing matrix
\item CP violation in charged gauge boson exchange (the usual CKM
mechanism)
\end{itemize}
It is clear that  particular 2HDMs show in general different
patterns for these CP violation sources. 
In this paper we analyse and compute the 2HDM predictions for the
flavour--diagonal 
CP--odd couplings of heavy fermions, both quarks and leptons, to the 
neutral gauge bosons $\gamma$ and $Z$. For
on--shell fermions and gauge bosons,
the corresponding $\bar f f\gamma$ and $\bar f f Z$ form factors are known
as the electric dipole moment (EDM) and weak electric dipole moment (WEDM) of
the fermion $f$ respectively. The presence of a nonvanishing dipole moment
of this kind is a signal of time reversal symmetry violation, and in our
framework, of CP violation. In general, the form factors are gauge invariant
quantities ---and can be contrasted with experiment--- only when the external
fermions and gauge bosons are on the mass shell. However, it can be seen that
in the 2HDM the one--loop predictions for the electric and weak electric
dipole form factors are still gauge invariant when the gauge bosons are
off shell.

We will present the
analysis for the CP--violating dipole form factors in the general 2HDM case,
and then apply the results to models which include discrete
symmetries and models in which the magnitude of the FCNC is related to the
masses of the involved fermions. As stated, the
fermion electric and weak electric dipole form factors are nonzero only at
high orders  in the SM, 
whereas 2HDM give rise in general to
nonvanishing contributions at the level of one loop. Therefore, they
represent good candidates for an  observation of CP--odd effects
arising from an extended
scalar sector. In the case of heavy fermions, the effects are
particularly important, since the new (scalar--mediated) contributions
are proportional to nonnegative powers of the fermion masses.

The paper is organized as follows: in section II the notation and the 
general 2HDM considered in this work are presented. The analytical and
numerical results for the CP--violating form factors are given in
section III, while section IV contains our conclusions. In the
appendix we quote some explicit expressions for the Feynman integrals
used in our analysis.

\section{Model}

As stated, we consider here a completely general 2HDM, allowing in
principle for the presence of tree--level FCNC. We adopt in the following
the notation introduced in Ref.\ \cite{wolf}, where the most general 
Higgs potential is parametrised as
\begin{eqnarray}
V(\phi_1,\phi_2) & = & -\mu_1^2\phi_1^\dagger\phi_1
-\mu_2^2\phi_2^\dagger\phi_2
-(\mu_{12}^2\phi_1^\dagger\phi_2+ {\rm h. c.}) \nonumber \\
& & +\lambda_1(\phi_1^\dagger\phi_1)^2
+\lambda_2(\phi_2^\dagger\phi_2)^2
+\lambda_3(\phi_1^\dagger\phi_1\,\phi_2^\dagger\phi_2)
+\lambda_4(\phi_1^\dagger\phi_2\,\phi_2^\dagger\phi_1) \nonumber \\
& & +\frac{1}{2}\left[\lambda_5(\phi_1^\dagger\phi_2)^2
+ {\rm h. c.}\right]
+\left[(\lambda_6 \phi_1^\dagger\phi_1+\lambda_7 \phi_2^\dagger\phi_2)
\,(\phi_1^\dagger\phi_2)+ {\rm h. c.}\right]\;,
\label{pot}
\end{eqnarray}
and the neutral scalars acquire vacuum expectation values
\begin{equation}
\langle\phi_1^0\rangle=\frac{v}{\sqrt{2}}e^{i\delta}\cos\beta
\,,\quad\quad
\langle\phi_2^0\rangle=\frac{v}{\sqrt{2}}\sin\beta   \;.
\label{vevs}
\end{equation}
In order to write the scalar--fermion couplings, it is convenient to
introduce a new basis for the scalars, namely
\begin{eqnarray}
\phi_1 & = & e^{i\delta}(\cos\beta\,\Phi_1+\sin\beta\,\Phi_2)
\nonumber\\
\phi_2 & = & \sin\beta\,\Phi_1 -\cos\beta\,\Phi_2
\end{eqnarray}
with
\begin{equation}
\Phi_1= \left( \begin{array}{c} G^+ \\
(v+H^0+iG^0)/\sqrt{2} \end{array} \right)\;,
\quad\quad\quad
\Phi_2= \left( \begin{array}{c} H^+ \\
(R+i I)/\sqrt{2} \end{array} \right) \;.
\end{equation}
It is easy to see that $H^\pm$ are physical charged scalar particles,
while $G^\pm$ and $G^0$ are the Goldstone bosons corresponding to the
spontaneous gauge symmetry breakdown. The remaining neutral scalars
$H^0$, $R$ and $I$ are not in general mass eigenstates.
In terms of these fields, the scalar--fermion couplings can be written as
\begin{equation}
L_Y=-\,(\sqrt{2} G_F)^{1/2}\, (L^{(nt)}+\sqrt{2}\, L^{(ch)}) \;,
\end{equation}
with
\begin{eqnarray}
L^{(nt)} & = & (\bar U_L M^u U_R +
\bar D_L M^d D_R +\bar L_L M^l L_R)\, H^0 + \nonumber \\
& & (\bar D_L \Gamma^d D_R + \bar L_L \Gamma^l L_R)\, (R+i I) +
\bar U_L \Gamma^u U_R\, (R-i I) + {\rm h. c.}\;,\nonumber\\
L^{(ch)} & = & \bar U_L V_{CKM} \Gamma^d D_R H^+ -
\bar D_L V_{CKM}^\dagger \Gamma^u U_R H^- + \bar N_L V_l\Gamma^l L_R H^+
+ {\rm h. c.} \;,
\label{ypot}
\end{eqnarray}
where we have used the definitions $U=(u,c,t)^T$, $D=(d,s,b)^T$,
$L=(e,\mu,\tau)^T$ and $N=(\nu_e,\nu_\mu,\nu_\tau)^T$. The couplings
 of the Goldstone bosons $G^0$ and  $G^\pm$ are
the standard ones. As we have checked that they do not
contribute to the form factors we are interested in, we have
not
written them explicitly in the  Lagrangian (\ref{ypot}).
 As
usual, $M^f$,
with $f=u,d,l$ stand for the quark and lepton diagonal mass matrices and
$V_{CKM}$ is the Cabibbo--Kobayashi--Maskawa matrix (the subindex $CKM$
will be omitted in the following to simplify the notation), whereas
$\Gamma^f$, $f=u,d,l$ are arbitrary $3\times 3$ complex matrices that arise
from the extended Yukawa couplings. It is useful to distinguish between
the diagonal and nondiagonal elements of $\Gamma^f$, defining~\cite{wolf}
\begin{equation}
\left(\Gamma^f\right)_{ij}=\left\{
\begin{array}{rl}
\xi_{f_i} m_{f_i}\,, & i=j \\
\mu_{ij}^f\,, & i\neq j
\end{array}
\right.\;,
\label{gammas}
\end{equation}
where $\xi_{f_i}$ and $\mu_{ij}^f$ are in general complex numbers. The
parameters $\mu_{ij}^f$ are responsible for the tree--level
flavour--changing neutral currents.

Nonzero phases in $\xi_{f_i}$ and $\mu_{ij}^f$
represent new sources of CP violation beyond the standard $\delta_{CKM}$
phase in $V$. In addition, a further source of CP violation arises from
the neutral scalar mixing: in the limit where CP is conserved, the
CP--even states $H^0$ and $R$ do not mix with $I$, which is CP--odd; however,
the nonhermitian terms in the Higgs potential can induce either explicit
or spontaneous CP violation [the latter, arising from the phase $\delta$
in (\ref{vevs})]. Then, in general, one expects the neutral scalars to
become mixed. The physical neutral mass eigenstates $H_i^0$ ($i=1,2,3$) can be
written as
\begin{equation}
H_i^0=\sum_{S=H^0,R,I} O_{Si} S \; ,
\label{higmix}
\end{equation}
being $O$ an orthogonal (real) matrix. It is clear that in general $H_i^0$
are not
eigenstates of CP.

\hfill

With the introduction of discrete symmetries to prevent FCNC, the above
general structure becomes simplified. By requiring the Yukawa couplings
to be invariant under the changes
\begin{eqnarray}
& \phi_1\rightarrow -\phi_1\;,\quad\quad \phi_2\rightarrow \phi_2\;, &
\nonumber \\
& D_{Ri}\rightarrow -D_{Ri}\;,\quad\quad L_{Ri}\rightarrow -L_{Ri}\;, &
\label{dsim}
\end{eqnarray}
together with $U_{Ri}\rightarrow -U_{Ri}$ ($U_{Ri}\rightarrow U_{Ri}$),
one obtains the so--called 2HDM I (II). The $3\times 3$ $\Gamma$ matrices
in (\ref{ypot}) are then given by
\begin{equation}
\Gamma^{d,l}=\tan\beta \;M^{d,l}\;,\quad\quad\Gamma^u=\left\{
\begin{array}{ll} \tan\beta\; M^u & \mbox{ (model I)} \\
-\cot\beta\; M^u & \mbox{ (model II)} \end{array} \right.\;\;.
\end{equation}
If the Higgs potential is also invariant under the transformations
(\ref{dsim}), all CP--violating terms in (\ref{pot}) turn out to be
forbidden. However, one can allow for a soft breakdown of the discrete
symmetry, retaining the CP violation only through the coupling with
$\mu_{12}^2$ in (\ref{pot}). The electric and weak electric dipole
form factors for the top quark have been analyzed within this scheme
in Ref.\ \cite{bern}. Notice that, in this case, the parameters
in (\ref{gammas}) satisfy Im$\xi_{f_i}=\mu_{ij}^f=0$, so that
the only source of CP violation beyond the SM is the mixing of
CP--even and CP--odd fields in (\ref{higmix}).

As commented above, some models propose specific relations between the
magnitude of the FCNC and the masses of the involved fermions. One
usual {\em Ansatz} is that proposed by Cheng and Sher~\cite{cs},
in which the matrix elements of $\Gamma^{u,d}$ are governed by the
order of magnitude of the fermion masses, obeying
\beq
\Gamma^{u,d}_{ij}=\lambda_{ij} \sqrt{m_i m_j}\;,
\label{chs}
\eeq
with $\lambda_{ij}$ not far from unity. In general, if the lightest
scalar masses are assumed to be in the region of a few hundreds GeV,
bounds from $\Delta F=2$ processes ($F=S,C,B$) constrain the couplings
$\lambda_{sd}$, $\lambda_{sb}$ and $\lambda_{uc}$ to be
$\alt 0.1$~\cite{atw}.
Nevertheless, present experimental information does not provide such kind
of constraints for $\lambda_{ct}$, which in principle is allowed to be
${\cal O}(1)$. Phenomenological consequences of having large $c$-$t$
flavour changing couplings have been studied recently by several
authors~\cite{varios}. We show below that the assumption of the
Cheng--Sher {\em Ansatz} of Eq.(\ref{chs}) with $\lambda_{ct}\sim 1$ leads to
a significant enhancement in the electric dipole form factors of the
$c$ quark.

\section{Analytical and numerical results for CP--violating dipole form
factors}

The most general Lorentz invariant matrix element for  the
interaction of a gauge boson $B$ with two on--shell fermions $f$, $\bar f$
can be written as
\begin{eqnarray}
\langle f(p_-) \bar{f}(p_+)|J_B^{\mu}(0)|B(q)\rangle
& & = i\ e\;\bar{u}_f(p_-) 
\left[F^{B,f}_V(q^2) \gamma_\mu+F^{B,f}_A(q^2)
\gamma_\mu\gamma_5\right.\nonumber\\
& &
\left.\hspace{-2cm}+\left(F^{B,f}_S(q^2)+F^{B,f}_{AN}(q^2)\gamma_5\right)
q_\mu
+\left(F^{B,f}_M(q^2)+F^{B,f}_E(q^2)\gamma_5\right) \sigma_{\mu\nu} q^\nu
\right] \; v_f(p_+)\;,
\label{ma}
\end{eqnarray}
where $q\equiv p_++p_-$, $e$ is the proton charge and the coefficients
$F_j^{B,f}$, so--called form factors, are in general
functions of $q^2$. At the tree--level, only the vector and axial vector
form factors can be different from zero in a gauge theory.

The last two coefficients, $F_M^{B,f}$ and $F_E^{B,f}$, are known as
magnetic (weak magnetic) and electric (weak electric) dipole
form factors when one considers the coupling with the gauge boson
$B=\gamma$ ($B=Z$). Both $F_M^{B,f}$ and $F_E^{B,f}$ are
chirality--flipping quantities.
Here we are interested in particular in $F_E^{B,f}$, which is CP--odd and
vanishing small  in the SM.
We
remark that within the SM all the above form factors $F_i^{B,f}(q^2)$ are
gauge independent only when the gauge boson $B$ is on shell. In this
case $F_E^{B,f}$ is called the electric ($B=\gamma$) or weak electric
($B=Z$) dipole moment of the fermion $f$.

We will analyse the values of $F_E^{B,f}(q^2)$ ($B=\gamma,Z$)
for heavy fermions in the 2HDM. Within these models one gets in general
contributions already at the one--loop level, and the form factors are
still gauge invariant when the gauge bosons are off--shell. 
Since these contributions are due 
to the exchange of neutral and charged Higgs bosons (which carry the
CP violation effects beyond the SM), and the Higgs--fermion couplings
are proportional to the corresponding fermion masses, only heavy fermions
are expected to provide significant effects.
We will concentrate in particular on the electric and weak electric
form factors for the $\tau$ lepton and the $t$, $b$ and $c$ quarks.

In the most general 2HDM, the relevant diagrams that contribute to
$F_E^{B,f}$ at one loop are shown in Fig.\ 1 [notice that those
in Fig.\ 1 (d) and (e) only contribute to the $Z \bar f f$ form factor].
Let us begin by quoting the results for $F_E^{Z,f}(q^2)$ being $f$ an
up--like quark. The contributions from the diagrams of Fig.\ 1 (a)--(e) are
given by
\begin{mathletters}
\begin{eqnarray}
(\mbox{a})\quad F_E^{Z,f}(q^2) & = &
\frac{2\sqrt{2}\, G_F}{\sw\cw} \, g_V^u
\nonumber \\
& & \times\sum_{f'=u,c,t} m_{f'} \sum_{j=1}^3
\bigg[m_{f}^2 O_{H^0j} \delta_{ff'}
(\im\xi_f O_{Rj}-\re\xi_f O_{Ij}) -
\re(\Gamma^u_{ff'}\Gamma^u_{f'f}) O_{Rj} O_{Ij} \nonumber \\
& & +\frac{1}{2}
\im(\Gamma^u_{ff'}\Gamma^u_{f'f})\, \left((O_{Rj})^2-(O_{Ij})^2\right)\bigg]
\;I^{\rm (I)}(m_f,m_{f'},m_{H^0_j},q^2)
\label{dza} \\
(\mbox{b})\quad F_E^{Z,f}(q^2) & = & \frac{2\sqrt{2}\, G_F}{\sw\cw} \, g_V^d
\nonumber \\
& & \times\sum_{f'=d,s,b} m_{f'}\,
\mbox{Im}\left[(\Gamma^{u\dagger} V)_{ff'}
(V \Gamma^d)^\ast_{ff'})\right]
I^{\rm (I)}(m_f,m_{f'},m_{H^\pm},q^2) \\
(\mbox{c})\quad F_E^{Z,f}(q^2) & = & \frac{2\sqrt{2}\, G_F}{\sw\cw} \, g^h
\nonumber \\
& & \times\sum_{f'=d,s,b} m_{f'}\,
\mbox{Im}\left[(\Gamma^{u\dagger} V)_{ff'}
(V \Gamma^d)^\ast_{ff'})\right]
I^{\rm (II)}(m_f,m_{f'},m_{H^\pm},q^2) \\
(\mbox{d})\quad F_E^{Z,f}(q^2) & = & \frac{2\sqrt{2}\, G_F}{\sw\cw}
\, g_V^u \, m_Z^2\, m_f \nonumber \\
& & \times\sum_{j=1}^3 (\im\xi_f O_{Rj}-\re\xi_f O_{Ij})\,
O_{H^0j}\;I^{\rm (III)}(m_f,m_{H^0_j},q^2) \label{dzd} \\
(\mbox{e})\quad F_E^{Z,f}(q^2) & = &
\frac{\sqrt{2}\, G_F}{2\sw\cw}\, m_f
\sum_{f'=u,c,t}\left(|\mu_{ff'}|^2-|\mu_{f'f}|^2\right)
\sum_{k<j} \left(O_{Rj} O_{Ik}-O_{Ij} O_{Rk}\right)
\nonumber \\
& & \times \left(O_{Rj} O_{Rk}+O_{Ij} O_{Ik}\right)\;
I^{\rm (IV)}(m_f,m_{f'},m_{H^0_j},m_{H^0_k},q^2)
\label{dze}
\end{eqnarray}
\label{dz}
\end{mathletters}
\hspace{-.27cm}where
\[
g_V^u=(\frac{1}{2}-\frac{4}{3}\swc)\,,\quad\quad\quad
g_V^d=(-\frac{1}{2}+\frac{2}{3}\swc)\,,\quad\quad\quad
g^h=(-\frac{1}{2} + \,\swc)\,,
\]
and the Feynman integrals $I^{\rm (i)}$ are quoted in Appendix A.
The gauge invariance of the form factors has been  explicitly checked.
For down--like quarks and leptons, the resulting expressions are similar
to those in Eqs.(\ref{dz}). In the case of the down quarks these are obtained
just by replacing
\begin{eqnarray}
\Gamma^u,\, g_V^u & \longleftrightarrow & \Gamma^d,\, g_V^d \nonumber\\
\sum_{f'=u,c,t} & \longleftrightarrow & \sum_{f'=d,s,b} \nonumber \\
O_{Ij} & \longrightarrow & -O_{Ij} \nonumber \\
g^h & \longrightarrow & -g^h \nonumber \\
V & \longrightarrow & V^\dagger
\end{eqnarray}
For the charged leptons, the diagrams in Fig.\ 1 (b) and (c) are zero in
the limit of vanishing neutrino masses. The remaining contributions can be
obtained from those in Eqs.\ (\ref{dz}) through the changes
\begin{eqnarray}
g_V^u & \longrightarrow & g_V^l \nonumber \\
\sum_{f'=u,c,t} & \longrightarrow & \sum_{f'=e,\mu,\tau} \nonumber \\
O_{Ij} & \longrightarrow & -O_{Ij}
\end{eqnarray}
where $g_V^l=-1/2+2\swc$.

Finally, the contributions to the electric dipole form
factors $F_E^{\gamma,f}$ are also obtained easily from the
corresponding expressions for $F_E^{Z,f}$. In this case the rule simply
consists in the replacements
\begin{equation}
\frac{g_V^f}{\sin\theta_W\cos\theta_W} \longrightarrow 2 Q_f\,,
\quad\quad\quad
\frac{g^h}{\sin\theta_W\cos\theta_W} \longrightarrow 1\;.
\end{equation}
for $f=u,d,l$. As stated, the diagrams in Fig.\ 1 (d) and (e) do not
contribute to $F_E^{\gamma,f}$.

Now we can make use of these results to evaluate the leading contributions
to the electric and weak electric dipole form factors for the $t$, $b$ and
$c$ quarks and the $\tau$ lepton. The final expressions can be written as
\begin{eqnarray}
F_E^{Z,t}(q^2) & = & \sum_{i=1}^3 \left(a_j^{Z,t} \alpha_j^t+
d_j^{Z,t}\gamma_j^t\right) \nonumber \\
F_E^{\gamma,t}(q^2) & = & \sum_{i=1}^3 a_j^{\gamma,t} \alpha_j^t \nonumber \\
F_E^{Z,b}(q^2) & = & (b+c)^{Z,b} \beta^b + \sum_{i=1}^3 d_j^{Z,b}\gamma_j^b
\nonumber \\
F_E^{\gamma,b}(q^2) & = & (b+c)^{\gamma,b} \beta^b +
\sum_{i=1}^3 a_j^{\gamma,b} \alpha_j^b \nonumber \\
F_E^{Z,\tau}(q^2) & = & \sum_{i=1}^3 d_j^{Z,\tau}\gamma_j^\tau
\nonumber \\
F_E^{\gamma,\tau}(q^2) & = & \sum_{i=1}^3 a_j^{\gamma,\tau} \alpha_j^\tau
\nonumber \\
F_E^{Z,c}(q^2) & = & \sum_{i=1}^3 \left({a'_j}^{Z,c} {\alpha'_j}^c+
d_j^{Z,c}\gamma_j^c\right)\nonumber \\
F_E^{\gamma,c}(q^2) & = & \sum_{i=1}^3 \left(a_j^{\gamma,c}\alpha_j^c+
{a'_j}^{\gamma,c} {\alpha'_j}^c \right)+ (b'+c')^{\gamma,c} {\beta'}^c
\label{ffact}
\end{eqnarray}
In this parametrisation, the dependence of $F_E^{B,f}$ with the
unknown quark mass--matrix parameters and Higgs--mixing angles has been
collected in the factors in Greek letters, whereas the coefficients
$k_j^{B,f}$, with $k=a$, $a'$,\ ...\ $d$, contain the global factors
including gauge boson couplings and fermion and $Z$ masses,
plus the Feynman integrals, which depend on $q^2$ and the masses of
the Higgs bosons. The
letters $a,b,c,d$ identify the diagram from which each contribution originates,
 according to the notation in Fig.\ 1
and Eqs.\ (\ref{dz}). We
have distinguished with primes the contributions that include tree--level
flavour--changing effects.

In Eqs.\ (\ref{ffact}) we have quoted only the dominant terms arising
from the expressions (\ref{dz}), hence the contributions proportional to
light fermion masses have been neglected. In addition, the contributions
from the diagram in Fig.\ 1 (e), which are proportional to flavour--changing
parameters, have been neglected in comparison to flavour--changing
terms arising from the diagram in Fig.\ 1 (a) [notice that (\ref{dze})
vanishes when the matrices $\Gamma^f$ are hermitian].

The explicit expressions for the factors in Greek letters can be easily
obtained from Eqs.\ (\ref{dz}). We find
\begin{mathletters}
\begin{eqnarray}
\alpha_j^f & = & O_{H^0j} \left[O_{Rj}\,{\rm Im}\xi_f
-\epsilon\, O_{Ij}\,{\rm Re}\xi_f\right] -\epsilon 
\,{\rm Re}(\xi_f^2) O_{Rj} O_{Ij}+
\frac{1}{2}{\rm Im}(\xi_f^2) \left(O_{Rj}^2-O_{Ij}^2\right) \label{gria}\\
\beta^b & = & - {\rm Im}(\xi_t\xi_b) \label{grib} \\
\gamma_j^f & = & O_{H^0j} \left(O_{Rj}\,{\rm Im}\xi_f
-\epsilon\, O_{Ij}\,{\rm Re}\xi_f\right) \label{gric} \\
{\alpha'}^c & = & -
\frac{{\rm Re}(\mu_{ct}\,\mu_{tc})}{m_c m_t}\, O_{Rj} O_{Ij} +\frac{1}{2}\;
\frac{{\rm Im}(\mu_{ct}\,\mu_{tc})}{m_c m_t}\,(O_{Rj}^2-O_{Ij}^2)
\label{grid} \\
{\beta'}^c & = & \sqrt{\frac{m_c}{m_t}}\;|V_{cb}|^2\,
{\rm Im}(\xi_c\xi_b)
+\sqrt{\frac{m_s}{m_b}}\; \frac{{\rm Im}(\mu_{tc}\mu_{sb}V_{cs}
V_{tb}^\ast)}{\sqrt{m_s m_c m_b m_t}}
+\frac{{\rm Im}(\mu_{tc}\xi_bV_{cb}V_{tb}^\ast)}{\sqrt{m_c m_t}}
\label{grie}
\end{eqnarray}
\label{grieg}
\end{mathletters}
\hspace{-.27cm}where $\epsilon=+1$ for $f=u,t$ and $\epsilon=-1$ for
$f=b,\tau$. If we
assume that $|\xi_f|$ is not very different from one for all fermions
(as is the case in most models for the quark mass matrices), all
parameters in Eqs.\ (\ref{gria}) -- (\ref{gric}) are expected to be
${\cal O}(1)$. Then, if no accidental cancellations occur,
the order of magnitude for the electric and weak electric dipole form
factors will be given by
the coefficients $k_j^{B,f}$ in Eqs.(\ref{ffact}). On the other hand, in the
case of the $c$ quark  we have to deal with contributions proportional
to ${\alpha'}^c$
and ${\beta'}^c$, which contain the flavour--changing parameters
$\mu_{ct}$, $\mu_{tc}$ and $\mu_{sb}$. These contributions depend on
the {\em Ansatz} chosen for the quark--mass matrix, and
can be very important due to the large top--quark mass.

To estimate the order of magnitude of the CP--violating form factors in
different 2HDM scenarios, we have calculated numerically the values of
the coefficients $k_j^{B,f}$ for different values of $q^2$ and the Higgs
masses. Our results are presented in Tables I--IV.
For the $b$, $c$, and $\tau$ form factors we have chosen
$\sqrt{q^2}=10$, 92, 170 and 500 GeV, corresponding to the
approximate centre--of--mass
energies in $B$-meson factories, LEP1, LEP2 and future $e^+e^-$ colliders
respectively. For the $t$ quark we have taken $\sqrt{q^2}=500$ and 1800 GeV,
the latter corresponding to the high--energy $\bar p p$ collider at
Fermilab. We have considered neutral and charged scalar--boson masses of
100 and 200 GeV. In general, the Feynman integrals are expected to be
suppressed if the masses of the involved scalars increase, so that the
sums in (\ref{ffact}) should be dominated by the contribution from the
lightest Higgs. Notice that in the limit where the neutral scalar sector
is degenerate in mass, the contributions from the diagrams Fig.\ 1 (a),
(d) and (e) to both the electric and weak electric dipole form factors
vanish owing to the orthogonality of the neutral--scalar mixing matrix~$O$.

The contributions of the different CP violation sources in the general
2HDM can be easily
read from Eqs.\ (\ref{grieg}). Let us come back to the classification
presented in the introduction: the CP violation in charged
and neutral flavour--conserving scalar exchange is contained in the
imaginary part of the parameters $\xi_f$, whereas the CP violation in neutral
flavour--changing scalar exchange is due to the imaginary part of $\mu_{ff'}$.
The CP violation in the scalar mixing matrix arises from the
mixing between the CP--odd scalar $I$ and the CP--even fields $H^0$ and $R$,
that means, the products $O_{H^0j}O_{Ij}$ and $O_{Rj}O_{Ij}$ in (\ref{grieg}).
Finally the $\delta_{CKM}$ CP--violating phase in the $V$ matrix only appears
together with tree--level flavour--changing parameters in Eq.\ (\ref{grie}).

In Table V, we quote the expected orders of magnitude of the
electric and weak electric form factors at $q^2=m_Z^2$ for the $\tau$
lepton and the $b$ and $c$ quarks, in both 2HDM I/II and Cheng--Sher--like
scenarios.
For the 2HDM I and II, which include discrete symmetries to prevent FCNC,
some of the contributions in (\ref{ffact}) vanish. As stated in the previous
section, in this case
Im$\xi_f=\mu_{ff'}=0$, hence $\beta^f={\alpha'}^f={\beta'}^f=0$, and all
the effect arises from the CP violation in the Higgs--mixing matrix.
On the contrary, in the quark--mixing
scheme proposed by Cheng and Sher, all terms in Eqs.(\ref{ffact}) have to
be considered if the $\lambda_{ij}$ parameters of Eq.\ (\ref{chs}) are
complex numbers of
order one. For the case of the $b$ and $c$ quarks, some of the terms that
vanish in the 2HDM I and II have relatively large coefficients, proportional
to the square of the top--quark mass. This shows up in the value of the
corresponding electric dipole form factors, where the predictions in the 
Cheng-Sher scheme are
about three orders of magnitude higher than in the 2HDM I/II, as
can be seen in Table V. In the
case of the weak electric dipole form factors the effect is hidden due to
the presence of other important contributions proportional to $m_Z^2$.

For the top quark the CP--violating form factors in the
general 2HDM are dominated by the contributions of diagrams (a) and (d)
in Fig.\ 1. These arise from flavour--conserving neutral Higgs exchange,
and do not vanish in general in 2HDM I and II. As stated in the previous
section, $F_E^{Z,t}$ and $F_E^{\gamma,t}$ have been analyzed previously for
these models in Ref.~\cite{bern}. The values for $F_E^{Z,t}(q^2)$ arising
from Eqs.\ (\ref{dza}) and (\ref{dzd}) and the equivalent
expressions for $F_E^{\gamma,t}(q^2)$ are in agreement with the results
obtained in that paper.

The weak electric dipole moments of the $\tau$ lepton and $b$ quark have
been also calculated recently within the Minimal Supersymmetric Standard
Model (MSSM)~\cite{susy}. The results are similar to those obtained
in our general 2HDM scheme: $|\mbox{Re}[F_E^{Z,\tau}(m_Z^2)]|\alt 0.3(12)
\times 10^{-21}$ $e\,$cm, $|\mbox{Re}[F_E^{Z,b}(m_Z^2)]|\alt 1.4(35)
\times 10^{-21}$ $e\,$cm. In Ref.~\cite{susy} the authors also quote the
values obtained for the top quark form factors at $\sqrt{q^2}=500$ GeV,
which yield approximately $|F_E^{Z,t}|\simeq
|F_E^{\gamma,t}|\simeq 10^{-19}$ $e\,$cm.
Taking the corresponding coefficients from Table I,
this order of magnitude agrees with our results.
Notice that in our analysis we have not considered possible corrections
arising from the running of the quark masses (we have used the running
masses $m_q$ in the $\overline{\rm MS}$ scheme, with $\mu=m_q$).
In any case, these effects would not modify the quoted orders of
magnitude for the values in Tables I--V.

\section{Conclusions}

We compute the CP--violating electric and weak electric dipole form
factors for heavy fermions in the framework of a completely general
2HDM. Despite of being one of the simplest extensions of the SM, this
model contains interesting new features, such as the presence of
various sources of CP violation beyond the standard CKM mechanism.
The CP--violating dipole form factors are vanishingly small in the SM,
thus they are good candidates to provide observable signals of new physics.
In the 2HDM, at one loop, they are found to be finite and gauge invariant
quantities, even when the involved $\gamma$ or $Z$ bosons are off--shell.

The effect of the different sources of CP violation on the form factors
for the $c$, $b$ and $t$ quarks and the $\tau$ lepton is shown
in Eqs.~(\ref{grieg}).
In particular, it is seen that some of the contributions vanish
for the so--called 2HDM I and II, which include discrete symmetries to
eliminate undesired FCNC. In these models the only remaining terms are
those involving the mixing between CP--even and CP--odd neutral scalars.
In the general case, however, these symmetries may be not present. If no
accidental cancellations occur, this would imply an enhancement of
three orders of magnitude in the electric dipole form factor of the $b$
quark with respect to the prediction of 2HDM I and II for energies in
the GeV to TeV range. On the other hand, we find that in the case of the
$c$ quark the electric dipole form
factor is strongly dependent on the presence of $c$-$t$ flavour--changing
effects. Assuming an up--quark mass matrix of the type proposed by
Cheng and Sher, with $\lambda_{uc}\sim {\cal O}(1)$, the values for
$F_E^{\gamma,c}$ for energies from 10 to 500 GeV can be two to four orders
of magnitude larger than those obtained in 2HDM I or II.
We conclude from this analysis that the study of CP--violating dipole form
factors can be useful to get information on the flavour mixing, and
offers an interesting possibility to distinguish between these and other
possible 2HDM scenarios.

\acknowledgements
We would like to thank A. Santamaria for
 valuable comments. 
D. G. D.\ has been supported by a grant from the Commission
of the European Communities, under the TMR programme (Contract
N$^\circ$ ERBFMBICT961548). This work has been funded by
CICYT (Spain) under grant No. AEN-96-1718, by AECI (Spain) and by 
CSIC (Uruguay).

\appendix

\section{Feynman integrals}

The integrals $I^{\rm (i)}$ introduced in Eq.\ (\ref{dz}) are defined
as follows:
\begin{eqnarray}
\re[I^{\rm (I)}(m_q,m_{q'},m_\phi,s)] & = & \frac{1}{16\pi^2}
\;\mbox{P.V.}\,\int_0^1 dx \int_0^{1-x} dy\, \frac{x}{f_1(x,y)}\;, \\
& & \hspace{-5cm} f_1(x,y)=m_\phi^2\, (1-x-y)+(m_{q'}^2-m_q^2)\,
(x+y)+ m_q^2\, (x+y)^2 - s\, x\, y \nonumber\\
\nonumber \\
\im[I^{\rm (I)}(m_q,m_{q'},m_\phi,s)] & = & \frac{1}{16\pi\, s}\;
\frac{\beta_{q'}}{\beta_q^2}\left\{1+\frac{1}{\beta_q\beta_{q'}}\left(
\frac{\beta_q^2-{\beta_{q'}}^2}{4}-\frac{m_\phi^2}{s}\right)\right.
\nonumber \\
& & \left.\times\log\left(\frac{s\,(\beta_q+\beta_{q'})^2
+4m_\phi^2}{s\,(\beta_q-\beta_{q'})^2+4m_\phi^2}\right)\right\}\,
\Theta(s-4m_{q'}^2) \label{im1} \\
\nonumber \\
\re[I^{\rm (II)}(m_q,m_{q'},m_\phi,s)] & = & \frac{1}{16\pi^2}\;\mbox{P.V.}\,
\int_0^1 dx \int_0^{1-x} dy\, \frac{1-2\,x}{f_2(x,y)}\;, \\
& & \hspace{-5cm}
f_2(x,y)=m_{q'}^2\, (1-x-y)+(m_\phi^2-m_q^2)\, (x+y)+ m_q^2\, (x+y)^2 -
s\, x\, y \nonumber \\
\nonumber \\
\im[I^{\rm (II)}(m_q,m_{q'},m_\phi,s)] & = & -\frac{1}{8\pi\, s}\;
\frac{\beta_\phi}{\beta_q^2}\left\{1-\frac{1}{\beta_q\beta_\phi}\left(
\frac{\beta_q^2+\beta_\phi^2}{4}+\frac{m_{q'}^2}{s}\right)\right.
\nonumber \\
& & \left.\times\log\left(\frac{s\,(\beta_q+\beta_\phi)^2
+4m_{q'}^2}{s\,(\beta_q-\beta_\phi)^2+4m_{q'}^2}\right)\right\}
\Theta(s-4m_\phi^2) \\
\nonumber \\
\re[I^{\rm (III)}(m_q,m_\phi,s)] & = & -\frac{1}{16\pi^2}\;\mbox{P.V.}\,
\int_0^1 dx \int_0^{1-x} dy\, \frac{y}{f_3(x,y)}\;, \\
& & \hspace{-5cm}
f_3(x,y)=m_q^2\, (1-x-y)^2+m_\phi^2\, x+ m_Z^2\, y - s\, x\, y
\nonumber \\
\nonumber \\
\im[I^{\rm (III)}(m_q,m_\phi,s)] & = & -\frac{1}{16\pi\, s}\;
\left\{\frac{b_Z}{\beta_q^2}+\frac{1}{2\beta_q}\left(
\frac{c-\beta_q^2}{\beta_q^2}+\frac{m_Z^2-m_\phi^2}{s}\right)\right.
\nonumber \\
& & \left.\times\log\left|\frac{c-b_Z\beta_q}{c+b_Z\beta_q}
\right|\,\right\} \Theta[s-(m_Z+m_\phi)^2] \label{im3} \\
\nonumber \\
\re[I^{\rm (IV)}(m_q,m_{q'},m_\phi,m_{\phi'},s)] & = &
-\frac{1}{16\pi^2}\;\mbox{P.V.}\,
\int_0^1 dx \int_0^{1-x} dy\, \frac{(x-y)\,(1-x-y)}{f_4(x,y)}\;, \\
& & \hspace{-5cm}
f_4(x,y)=(m_{q'}^2-m_q^2)\, (1-x-y)+m_q^2\, (1-x-y)^2
+m_\phi^2\, x+m_{\phi'}^2\, y - s\, x\, y
\nonumber \\
\nonumber \\
\im[I^{\rm (IV)}(m_q,m_{q'},m_\phi,m_{\phi'},s)] & = &
-\frac{1}{16\pi}\,\frac{(m_\phi-m_{\phi'})}{s^2}
\nonumber \\
& & \times\left\{\frac{2\,b_{\phi'}}{\beta_q^2}+\frac{d}{\beta_q}
\log\left|\frac{d-b_{\phi'}\beta_q}{d+b_{\phi'}\beta_q}
\right|\,\right\} \Theta[s-(m_\phi+m_{\phi'})^2]
\end{eqnarray}
with
\begin{eqnarray}
\beta_a & = & (1-4m_a^2/s)^{1/2}\;,\quad a=q,q',\phi \nonumber \\
b_a & = & [(1-(m_\phi+m_a)^2/s]^{1/2}\,
[1-(m_\phi-m_a)^2/s]^{1/2}\;,\quad a=Z,\phi' \nonumber \\
c & = & 1-(m_\phi^2+m_Z^2)/s \nonumber \\
d & = & 1-(m_\phi^2+m_{\phi'}^2)/s -2\,(m_q^2-m_{q'}^2)/s
\end{eqnarray}

In these expressions we assume $s > 4m_q^2$, which is
valid for all the numerical estimations that are presented in this paper.
The analytical results in (\ref{im1}) and (\ref{im3}) agree with those
presented in Ref.~\cite{bern} in the limit $\beta_{q'}=\beta_q$.

\begin{figure}[htbp]
\begin{center}
\epsfig{file=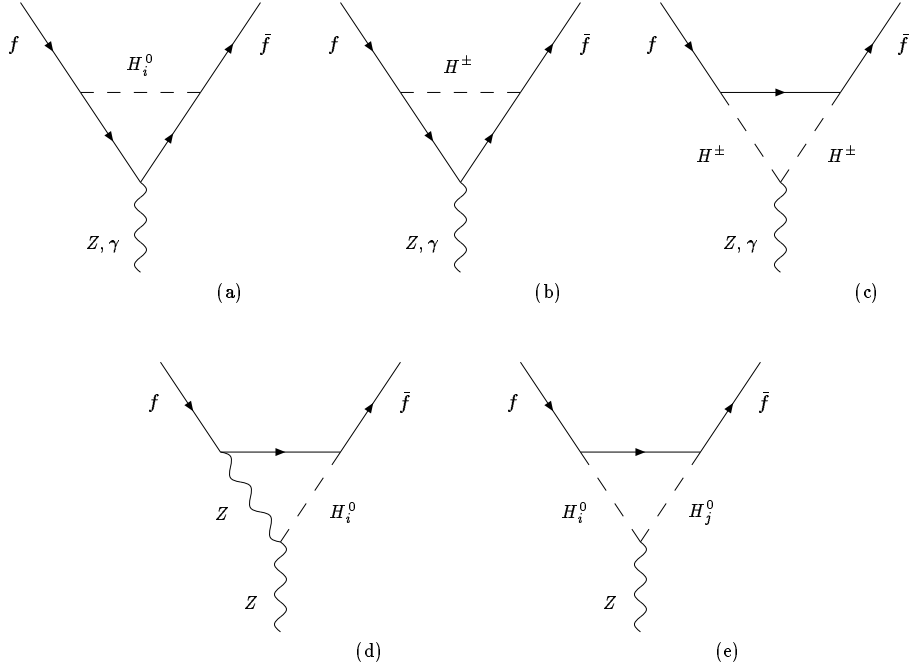}
\end{center}
\caption{One-loop contributions to the form factors $F_E^Z(q^2)$
and $F_E^\gamma(q^2)$ in the 2HDM.}
\label{fig1}
\end{figure}

\begin{table} 
\caption{Dominant coefficients for $F_E^{Z,t}(s)$ and $F_E^{\gamma,t}(s)$
for different values of $s$ and the Higgs mass $m_H$. All values are in
units of $e\,$cm.}
\begin{tabular}{l||r|r|r|r}
\hspace{0cm} & \mco {2}{c|}{$\sqrt{s} = 500$ GeV} &
\mco {2}{c}{$\sqrt{s} = 1800$ GeV} \\ 
\hspace{0cm} & \mco {1}{c}{$m_H = 100$ GeV} &
\mco {1}{c|}{$m_H = 200$ GeV} & \mco {1}{c}{$m_H = 100$ GeV} &
\mco {1}{c}{$m_H = 200$ GeV} \\ 
\hline
$a^{Z,t}$ &
$(-0.1+1.4i)\times 10^{-19}$ &
$(0.2+1.0i)\times 10^{-19}$ &
$(-1.2+1.0i)\times 10^{-20}$ &
$(-1.0+1.0i)\times 10^{-20}$ \\
$d^{\gamma,t}$ &
$(1.0-2.8i)\times 10^{-20}$ &
$(0.8-3.5i)\times 10^{-20}$ &
$(3.1-2.6i)\times 10^{-21}$ &
$(3.0-2.6i)\times 10^{-21}$ \\ \hline
$a^{\gamma,t}$ &
$(-0.3+4.1i)\times 10^{-19}$ &
$(0.6+2.9i)\times 10^{-19}$ &
$(-3.4+2.9i)\times 10^{-20}$ &
$(-2.9+2.7i)\times 10^{-20}$
\end{tabular}
\end{table}

\begin{table}
\caption{Dominant coefficients for $F_E^{Z,b}(s)$ and $F_E^{\gamma,b}(s)$
for $m_H=100$ and 200 GeV and different values of $s$. All values are in
units of $e\,$cm.} 
\begin{tabular}{l||c|c|c|r}
\hspace{0cm} & $\sqrt{s} = 10$ GeV &
$\sqrt{s} = m_Z$ & $\sqrt{s} = 170$ GeV & \mco {1}{c}{$\sqrt{s} = 500$ GeV} \\ 
\hline \hline
\hspace{0cm} & \mco{4}{c}{$m_H=100$ GeV} \\ \hline
$(b+c)^{Z,b}$ &
$5.0\times 10^{-21}$ &
$5.1\times 10^{-21}$ &
$5.7\times 10^{-21}$ &
$(2.8+5.9 i)\times 10^{-21}$ \\ 
$d^{Z,b}$ &
$3.6\times 10^{-21}$ &
$4.0\times 10^{-21}$ &
$6.3\times 10^{-21}$ &
$(-1.1+1.8 i)\times 10^{-21}$ \\ \hline
$a^{\gamma,b}$ &
$(-0.8-\!0.2i)\times 10^{-22}$ &
$(-0.7-2.6i)\times 10^{-23}$ &
$(0.2-1.4i)\times 10^{-23}$ &
$(2.1-2.7i)\times 10^{-24}$ \\
$(b+c)^{\gamma,b}$ &
$1.0\times 10^{-20}$ &
$1.0\times 10^{-20}$ &
$1.1\times 10^{-20}$ &
$(0.5+1.3 i)\times 10^{-20}$ \\ \hline \hline
\hspace{0cm} & \mco{4}{c}{$m_H=200$ GeV} \\ \hline
$(b+c)^{Z,b}$ &
$3.0\times 10^{-21}$ &
$3.1\times 10^{-21}$ &
$3.3\times 10^{-21}$ &
$(3.4+3.2 i)\times 10^{-21}$ \\ 
$d^{Z,b}$ &
$2.0\times 10^{-21}$ &
$2.1\times 10^{-21}$ &
$2.4\times 10^{-21}$ &
$(-0.5+2.5 i)\times 10^{-21}$ \\ \hline
$a^{\gamma,b}$ &
$(-2.3-\!0.4i)\times 10^{-23}$ &
$(-0.6-0.9i)\times 10^{-23}$ &
$(-0.2-0.7i)\times 10^{-23}$ &
$(0.8-2.2i)\times 10^{-24}$ \\
$(b+c)^{\gamma,b}$ &
$6.5\times 10^{-21}$ &
$6.6\times 10^{-21}$ &
$7.0\times 10^{-21}$ &
$(6.5+7.8 i)\times 10^{-21}$
\end{tabular} 
\label{tabla2}
\end{table} 
 
\begin{table}
\centering
\caption{Dominant coefficients for $F_E^{Z,\tau}(s)$ and $F_E^{\gamma,\tau}(s)$
for $m_H=100$ and 200 GeV and different values of $s$. All values are in
units of $e\,$cm.} 
\begin{tabular}{l||c|c|c|r}
\hspace{0cm} & $\sqrt{s} = 10$ GeV & $\sqrt{s} = m_Z$ &
$\sqrt{s} = 170$ GeV & \mco {1}{c}{$\sqrt{s} = 500$ GeV} \\ 
\hline \hline
\hspace{0cm} & \mco{4}{c}{$m_H=100$ GeV} \\ \hline
$d^{Z,\tau}$ &
$1.6\times 10^{-22}$ &
$1.8\times 10^{-22}$ &
$2.9\times 10^{-22}$ &
$(-0.5+0.9 i)\times 10^{-22}$ \\ \hline
$a^{\gamma,\tau}$ &
$(-1.2-0.7i)\times 10^{-23}$ &
$(-1.0-4.8i)\times 10^{-24}$ &
$(0.4-2.7i)\times 10^{-24}$ &
$(0.4-0.5i)\times 10^{-24}$ \\
\hline \hline
\hspace{0cm} & \mco{4}{c}{$m_H=200$ GeV} \\ \hline
$d^{Z,\tau}$ &
$0.9\times 10^{-22}$ &
$0.9\times 10^{-22}$ &
$1.1\times 10^{-22}$ &
$(-0.2+1.1 i)\times 10^{-22}$ \\ \hline
$a^{\gamma,\tau}$ &
$(-0.4-0.2i)\times 10^{-23}$ &
$(-1.1-1.6i)\times 10^{-24}$ &
$(0.4-1.2i)\times 10^{-24}$ &
$(0.1-0.4i)\times 10^{-25}$
\end{tabular} 
\end{table} 

\begin{table}
\caption{Dominant coefficients for $F_E^{Z,c}(s)$ and $F_E^{\gamma,c}(s)$
for $m_H=100$ and 200 GeV and different values of $s$. All values are in
units of $e\,$cm.}
\begin{tabular}{l||c|c|c|r}
\hspace{0cm} & $\sqrt{s} = 10$ GeV &
$\sqrt{s} = m_Z$ & $\sqrt{s} = 170$ GeV & \mco {1}{c}{$\sqrt{s} = 500$ GeV} \\ 
\hline \hline
\hspace{0cm} & \mco{4}{c}{$m_H=100$ GeV} \\ \hline
${a'}^{Z,c}$ &
$0.6\times 10^{-21}$ &
$0.6\times 10^{-21}$ &
$0.7\times 10^{-21}$ &
$(0.2+1.0 i)\times 10^{-21}$ \\ 
$d^{Z,c}$ &
$-0.7\times 10^{-21}$ &
$-0.7\times 10^{-21}$ &
$-1.2\times 10^{-21}$ &
$(2.0-3.5 i)\times 10^{-22}$ \\ \hline
$a^{\gamma,c}$ &
$(5.9+2.8i)\times 10^{-24}$ &
$(0.5+1.9i)\times 10^{-24}$ &
$(-0.1+1.1i)\times 10^{-24}$ &
$(-1.6+2.0i)\times 10^{-25}$ \\
${a'}^{\gamma,c}$ &
$1.7\times 10^{-21}$ &
$1.8\times 10^{-21}$ &
$1.9\times 10^{-21}$ &
$(.6+2.9 i)\times 10^{-21}$ \\
$(b'+c')^{\gamma,c}$ &
$(5.3+1.0i)\times 10^{-24}$ &
$(1.0+1.6i)\times 10^{-24}$ &
$(0.6+0.9i)\times 10^{-24}$ &
$(0.5+7.5i)\times 10^{-25}$ \\
\hline \hline
\hspace{0cm} & \mco{4}{c}{$m_H=200$ GeV} \\ \hline
${a'}^{Z,c}$ &
$4.4\times 10^{-22}$ &
$4.6\times 10^{-22}$ &
$4.9\times 10^{-22}$ &
$(2.5+7.2 i)\times 10^{-22}$ \\ 
$d^{Z,c}$ &
$-3.7\times 10^{-22}$ &
$-3.8\times 10^{-22}$ &
$-4.4\times 10^{-22}$ &
$(1.0-4.6 i)\times 10^{-22}$ \\ \hline
$a^{\gamma,c}$ &
$(1.6+0.7i)\times 10^{-24}$ &
$(4.3+6.4i)\times 10^{-25}$ &
$(1.4+5.0i)\times 10^{-25}$ &
$(-0.6+1.6i)\times 10^{-25}$ \\
${a'}^{\gamma,c}$ &
$1.3\times 10^{-21}$ &
$1.3\times 10^{-21}$ &
$1.4\times 10^{-21}$ &
$(0.7+2.1 i)\times 10^{-21}$ \\
$(b'+c')^{\gamma,c}$ &
$(1.6+0.3i)\times 10^{-24}$ &
$(0.5+0.5i)\times 10^{-24}$ &
$(2.7+4.1i)\times 10^{-25}$ &
$(-0.2+1.4i)\times 10^{-25}$
\end{tabular} 
\label{tabla4}
\end{table} 
 
\begin{table}
\caption{Expected order of magnitude for electric and weak electric form
factors at $q^2=m_Z^2$ for $\tau$ lepton and $b$ and $c$ quarks
in 2HDM I/II and Cheng--Sher--like scenarios.
Values are in units of $e\,$cm.}
\begin{tabular}{l||c|c|c|c|c|c}
\hspace{0cm} & $F^{Z,\tau}_E(m_Z^2)$ & $F^{\gamma,\tau}_E(m_Z^2)$ &
$F^{Z,b}_E(m_Z^2)$ & $F^{\gamma,b}_E(m_Z^2)$ &
$F^{Z,c}_E(m_Z^2)$ & $F^{\gamma,c}_E(m_Z^2)$ \\
\hline
2HDM I/II & $10^{-22}$ & $10^{-24}$ & $10^{-21}$ & $10^{-23}$ & $10^{-21}$
& $10^{-24}$ \\
Cheng--Sher & $10^{-22}$ & $10^{-24}$ & $10^{-20}$-$10^{-21}$
& $10^{-20}$ & $10^{-21}$-$10^{-21}|\lambda_{ct}|^2$
& $10^{-21}|\lambda_{ct}|^2$
\end{tabular}
\end{table}


\begin{references} 

\bibitem{ale} W. Bernreuther \etal , Z. Phys. C {\bf 52} (1994) 567.

W. Bernreuther, O. Nachtmann and P. Overmann, Phys. Rev. D {\bf 48} (1993) 78.

\bibitem{nos1} J. Bernab\'eu, G. A. Gonz\'alez-Sprinberg and J. Vidal, Phys. 
Lett. B {\bf 326} (1994) 168.

\bibitem{exp} M. Acciarri \etal , Phys. Lett. B {\bf 426} (1998) 207.

K. Ackerstaff \etal , Z. Phys. C {\bf 74}, (1997) 403.

D. Buskulic \etal , Phys. Lett. B {\bf 346} (1995) 371.

\bibitem{bern} W. Bernreuther, T. Schr\"oder and T. N. Pham,
Phys.  Lett. {\bf B279} (1992) 389.

\bibitem{susy} W.\ Hollik, J.\ I.\ Illana, S.\ Rigolin and D.\ St\"ockinger,
Phys.\ Lett.\ B {\bf 425} (1998) 322.

W.\ Hollik, J.\ I.\ Illana, S.\ Rigolin,
C.\ Schappacher and D.\ St\"ockinger, preprint DESY 98--195,
KA--TP--20--1998, December 1998, {\tt hep-ph/9812298}.

\bibitem{nel} C.\ A.\ Nelson, Phys. Rev. D {\bf 41} (1990)  2805.

\bibitem{sant} T.\ Barklow, contribution to TAU98, Santander, Spain,
Sept. 1998.

\bibitem{glawei} S.\ L.\ Glashow and S.\ Weinberg,
Phys.\ Rev.\ D {\bf 15} (1977) 1958.

\bibitem{mod} J.\ Liu and L.\ Wolfenstein, Nucl.\ Phys.\
{\bf B289} (1987) 1.

\bibitem{cs} T.\ P.\ Cheng and M.\ Sher, Phys.\ Rev.\ D {\bf 35}
(1987) 3484.

\bibitem{afs} A.\ Antaramian, L.\ J.\ Hall and A.\ Ra\v{s}in, Phys.\
Rev.\ Lett.\ {\bf 69} (1992) 1871.

L.\ J.\ Hall and S.\ Weinberg, Phys.\ 
Rev.\ D {\bf 48} (1993) 979.

\bibitem{wolf} Y.-L.\ Wu and L.\ Wolfenstein, Phys.\ Rev.\ Lett.\
{\bf 73} (1994) 1762. 

\bibitem{atw} D.\ Atwood, L.\ Reina and A.\ Soni, Phys.\ Rev.\ D
{\bf 55} (1997) 3156.

\bibitem{varios} T.\ Han, K.\ Whisnant, B.\ -L.\ Young and X.\ Zhang,
Phys.\ Lett.\ B {\bf 385} (1996) 311.

S.\ Bar-Shalom, G.\ Eilam and A.\ Soni, Phys.\ Rev.\
Lett.\ {\bf 79} (1997) 1217; Phys.\ Rev.\ D {\bf 57} (1998) 2957.

W.\ -S.\ Hou, G.\ -L.\ Lin and C.\ -Y.\ Ma, Phys.\ Rev.\ D {\bf 56}
(1997) 7434.

\end{references}
\end{document}